\documentclass[journal]{IEEEtran}

\usepackage[]{graphicx}    
\newcommand{\ignore}[1]{}  
\usepackage{color}
\usepackage[table]{xcolor}
\usepackage{cite}
\usepackage{epsfig,psfrag}
\usepackage{amssymb,amsmath,amsfonts}
\usepackage{bm}
\usepackage{tabularx}
\usepackage{multirow}
\usepackage[font=scriptsize]{subcaption}
\usepackage{setspace}
\usepackage{amsthm}
\usepackage{tikz}
\usepackage{pst-node}
\usepackage{acronym}
\usepackage[yyyymmdd,hhmmss]{datetime}
\usepackage{notation}
\usepackage{diagbox}
\usepackage[normalem]{ulem}
\usetikzlibrary{arrows,backgrounds,calc,positioning,shapes,shadows,matrix}

\providecommand{\ist}{\hspace*{.3mm}}
\providecommand{\rmv}{\hspace*{-.3mm}}

\providecommand{\nn}{\nonumber}
\newcommand{\T}{\mathrm{T}}

\newcolumntype{L}[1]{>{\raggedright\arraybackslash}p{#1}}
\newcolumntype{C}[1]{>{\centering\arraybackslash}p{#1}}
\newcolumntype{R}[1]{>{\raggedleft\arraybackslash}p{#1}}

\acrodef{gmm}[GMM]{Gaussian mixture model}
\acrodef{da}[DA]{data association}
\acrodef{mmse}[MMSE]{minimum mean-square error}
\acrodef{ps}[PS]{potential source}
\acrodef{po}[PO]{potential object}
\acrodef{pmf}[PMF]{probability mass function}
\acrodef{pdf}[PDF]{probability density function}
\acrodef{iid}[iid]{independent and identically distributed}
\acrodef{rmse}[RMSE]{root-mean-squared error}
\acrodef{ospa}[OSPA]{optimal sub-pattern assignment}
\acrodef{bp}[BP]{belief propagation}
\acrodef{bpf}[BPF]{bootstrap particle filter}
\acrodef{upf}[UPF]{unscented particle filter}
\acrodef{pde}[PDE]{partial differential equation}
\acrodef{sde}[SDE]{stochastic differential equation}
\acrodef{ode}[ODE]{ordinary differential equation}
\acrodef{edh}[EDH]{exact Daum and Huang}
\acrodef{ledh}[LEDH]{localized exact Duam and Huang}
\acrodef{pfpf}[PFPF]{particle flow particle filter}
\acrodef{mcmc}[MCMC]{Markov Chain Monte Carlo}
\acrodef{smc}[SMC]{sequential Monte Carlo}
\acrodef{map}[MAP]{maximum a posteriori}
\acrodef{tdoa}[TDOA]{time-difference of arrival}
\acrodef{pfl}[PFL]{particle flow}
\acrodef{mot}[MOT]{multiobject tracking}
\acrodef{pda}[PDA]{probabilistic data association}
\acrodef{jpda}[JPDA]{Joint \ac{pda}}
\acrodef{phd}[PHD]{probability hypothesis density}
\acrodef{cphd}[CPHD]{cardinalized \ac{phd}}
\acrodef{mht}[MHT]{multi-hypothesis tracking}
\acrodef{slam}[SLAM]{simultaneous localization and mapping}
\acrodef{iid}[iid]{independent and identically distributed}
\acrodef{rfs}[RFS]{random finite sets}
\acrodef{ospa}[OSPA]{optimal sub-pattern assignment}
\acrodef{mospa}[MOSPA]{mean \ac{ospa}}
\acrodef{snr}[SNR]{signal-to-noise ratio}
\acrodef{roi}[ROI]{region of interest}
\acrodef{is}[IS]{importance sampling}

\definecolor{temporalgreen}{RGB}{0,128,0}
\definecolor{spatialred}{RGB}{255,0,0}
\definecolor{temporalblue}{RGB}{0,0,205}

\allowdisplaybreaks
\newcommand{\paperTitle}{Importance Sampling With Stochastic Particle Flow and Diffusion Optimization}
\newcommand{\paperTitleMarkboth}{Meyer, Etzlinger, Liu, Hlawatsch, and Win: A Scalable Algorithm for
Network Localization and Synchronization}

\begin{document}
\title{\paperTitle}
\author{Wenyu Zhang, Mohammad J.~Khojasteh, Nikolay A.~Atanasov, and Florian Meyer\\[-6mm]

\thanks{The material presented in this work was supported by the Qualcomm Innovation Fellowship No. 492866 and by the Office of Naval Research under Grant  N00014-23-1-2284.}

\thanks{W.~Zhang, N.~A.~Atanasov, and F.~Meyer are with the  University of California San Diego, La Jolla, CA, USA (e-mail: \texttt{wez078@ucsd.edu, natanasov@ucsd.edu, flmeyer} \texttt{@ucsd.edu}).}

\thanks{M.~J.~Khojasteh is with Rochester Institute of Technology, Rochester, NY, USA (e-mail: \texttt{mjkeme} \texttt{@rit.edu}).}
\vspace{0mm}
}

\maketitle

\begin{abstract} 
\Ac{pfl} is an effective method for overcoming particle degeneracy, the main limitation of particle filtering. In \ac{pfl}, particles are migrated towards regions of high likelihood based on the solution of a partial differential equation. Recently proposed stochastic \ac{pfl} introduces a diffusion term in the \ac{ode} that describes particle motion. This diffusion term reduces the stiffness of the \ac{ode} and makes it possible to perform \ac{pfl} with a lower number of numerical integration steps compared to traditional deterministic \ac{pfl}. In this work, we introduce a general approach to perform \ac{is} based on stochastic \ac{pfl}. Our method makes it possible to evaluate a ``flow-induced'' proposal \ac{pdf} after the parameters of a \ac{gmm} have been migrated by stochastic \ac{pfl}. Compared to conventional stochastic \ac{pfl}, the resulting processing step is asymptotically optimal. Within our method, it is possible to optimize the diffusion matrix that describes the diffusion term of the \ac{ode} to improve the accuracy-computational complexity tradeoff. Our simulation results in a highly nonlinear 3-D source localization scenario showcase a reduced stiffness of the \ac{ode} and an improved estimating accuracy compared to state-of-the-art deterministic and stochastic \ac{pfl}\vspace{-5.5mm}.
\end{abstract}

\section{Introduction}
The particle filter is probably the most widely used method for nonlinear sequential Bayesian estimation \cite{AruMasGorCla:02}. It can provide an asymptotically optimal approximation of posterior \acp{pdf} with complex shapes. The key operation performed in the update step of the particle filter is \ac{is}. Here, particles are first sampled from an arbitrary proposal \ac{pdf} and then weighted based on the likelihood function.  The particle filter is known to suffer from particle degeneracy in higher dimensional problems \cite{BicLiBen:B08}, i.e., due to the curse of dimensionality, too few particles have a significant weight after the update step.

To overcome particle degeneracy, \ac{pfl} \cite{DuaHua:07,DuaHua:09} migrates particles sampled from a predicted/prior \ac{pdf} to regions of high likelihood. By using a homotopy function, that describes the transition of the predicted/prior \ac{pdf} to the posterior \ac{pdf}, in the Fokker-Planck equation, one can either derive an \ac{ode} or \ac{sde} to describe the motion of particles. The resulting deterministic or stochastic \acp{pfl} have a  zero or non-zero \textit{diffusion} term, respectively. In particular, the \ac{edh} flow \cite{DuaHua:10} and the Gromov's flow \cite{DuaHuaNou:18} are popular deterministic and stochastic flows with closed-form solutions. Stochastic flows tend to require a lower number of \ac{pfl} steps due to improved transient dynamics, i.e., a reduced stiffness of the underlying \ac{ode}, and thus lead to a reduced overall computational complexity.  As proposed and demonstrated recently \cite{DaiDau:22}, in stochastic \ac{pfl} it is possible to optimize the diffusion term to further reduce stiffness and thus computational complexity. Although it has been demonstrated that \ac{pfl} can overcome particle degeneracy in a variety of nonlinear and high-dimensional problems \cite{DuaHua:07,DuaHua:09,DuaHua:10,DuaHuaNou:18}, it has no asymptotical optimality guarantee \cite{LiCoates:17}.

The \ac{pfpf}\cite{LiCoates:17} embeds \ac{pfl} into \ac{is} by introducing a flow-induced proposal \ac{pdf}. The evaluation of this proposal at the migrated particles requires an invertible mapping from the predicted/prior \ac{pdf} at particles before the flow and the \ac{pdf} represented by the particles after the flow. Invertible deterministic \ac{pfl} was recently combined with a \ac{gmm} and used within a belief propagation framework for the 3-D tracking of an unknown number of sources in the presence of data association uncertainty \cite{ZhaMey:21,ZhaMey:J24}. This approach has recently been used in the context of marine mammal research \cite{JanMeySny:J23}. However, this invertible mapping is limited to deterministic flows like the \ac{edh} \cite{BunGod:16}. For stochastic \ac{pfl}, in \cite{BunGod:16} particles are drawn at each flow step from a different proposal \ac{pdf}, which leads to a computationally expensive weights computation for \ac{is}. The work in \cite{LiZhaoCoates:15, PalCoa:19} utilizes auxiliary variables and their filtering for \ac{is} with embedded \ac{pfl}. In particular, \cite{PalCoa:19} extended this framework to the stochastic Gromov's flow. Although the method in \cite{PalCoa:19} is computationally less expensive than \cite{BunGod:16}, it has twice computational complexity of \ac{pfl} due to an additional auxiliary variable for each particle. Furthermore, it relies on heuristic and suboptimal covariance matrix selections  \cite{PalCoa:19}  and does not consider the most recent stochastic flows with optimized diffusion terms.

In this work, we introduce an approach that combines \ac{is} with general stochastic flows that include optimizable diffusion terms. The resulting \ac{is} framework is combined with a \ac{gmm} and used within a belief propagation framework for the detection and localization of an unknown number of sources in 3-D.  The main contributions of our work are summarized as follows.
\begin{itemize}
    \item We develop efficient \ac{is} where the parameters of a \ac{gmm} are migrated by stochastic \acp{pfl} that rely on an \ac{sde} with an optimizable diffusion term.
    \item We evaluate our method in a challenging 3-D multi-source localization problem and demonstrate significant improvements compared to state-of-the-art methods.
\end{itemize}

This paper advances over the preliminary account of our method provided in the conference publication \cite{ZhaKhoMey:24} by (i) extending the approach to general stochastic flows that include optimizable diffusion terms and  (ii) introducing additional and more extensive numerical results.

\section{Stochastic \ac{pfl}}
\vspace{-1.5mm}

Consider a random state to be estimated, $\V{x}\in \mathbb{R}^N$, in a Bayesian setting. \ac{pfl} \cite{DuaHua:07,DuaHua:09} establishes a continuous mapping w.r.t. to pseudo-time $\lambda\in [0,1]$, to migrate particles sampled from the the prior \ac{pdf} $\V{x}_0\sim f(\V{x})$ such that they represent the posterior \ac{pdf} $\V{x}_1\sim f(\V{x}|\V{z})$.  

Let $f(\V{x})$ be the prior \ac{pdf} and $h(\boldsymbol{x}) \rmv= \rmv f(\boldsymbol{z}|\boldsymbol{x})$ be the likelihood function, where $\boldsymbol{z}$ is observed and thus fixed. Following Bayes' rule, a log-homotopy function \cite{DuaHua:07,DuaHua:09}  can be introduced as $\phi(\boldsymbol{x},\lambda)=\log f(\boldsymbol{x})+\lambda\log h(\boldsymbol{x})$. By using this log-homotopy function in the Fokker-Plank equation and modeling the dynamics of the particles by an \ac{sde}\vspace{0mm}, i.e., 
\begin{equation}
    \mathrm{d} \V{x} = \V{\zeta}(\V{x}, \lambda) \mathrm{d}\lambda + \sqrt{\M{Q}(\lambda)}\mathrm{d} \V{w},
    \label{eq:stochasticFlow}
    \vspace{-0mm}
\end{equation}
stochastic \acp{pfl} can be derived \cite{DuaHuaNou:18}. (Note that modeling using an \ac{ode}, i.e., $\mathrm{d} \V{x} = \V{\zeta}(\V{x}, \lambda) \mathrm{d}\lambda$, leads to deterministic flows.)


In \eqref{eq:stochasticFlow}, $\V{\zeta}(\V{x}, \lambda)  \in \mathbb{R}^N$ is the drift vector, $\M{Q}(\lambda)\in \mathbb{R}^{N\times N}$ is the diffusion matrix, and $\V{w}\sim \mathcal{N}(\V{0},\M{I})$ is zero-mean Gaussian noise with covariance matrix $\M{I}$. Since the drift is both time and state-dependent, directly integrating $\lambda$ from 0 to 1 analytically is infeasible. Thus, the Euler-Maruyama method \cite{KloPla:B92} is commonly used for numerical integration. Here, particle migration is performed by evaluating $\V{\zeta}(\boldsymbol{x},\lambda)$ at $N_\lambda$ discrete values of $\lambda$, i.e., $0\rmv=\rmv\lambda_0\rmv<\rmv\lambda_1\rmv<\rmv...\rmv<\rmv\lambda_{N_\lambda}=1$. First, $N_\mathrm{s}$ particles $\big\{\V{x}_{0}^{(i)}\big\}_{i=1}^{N_\mathrm{s}}\! =\! \big\{\V{x}_{\lambda_{0}}^{(i)}\}_{i=1}^{N_\mathrm{s}}$ are drawn from $f({\boldsymbol{x}})$. Next, each particle $i\rmv\in\rmv \{1,\dots,N_\mathrm{s}\}$ is migrated sequentially across the discrete pseudo time steps $l \rmv\in\rmv\{1,\dots,N_{\lambda}\}$,\vspace{1.2mm} i.e.,
\begin{equation}
\boldsymbol{x}^{(i)}_{\lambda_l} = \boldsymbol{x}^{(i)}_{\lambda_{l-1}} + \V{\zeta}_{\mathrm{s}}(\boldsymbol{x}^{(i)}_{\lambda_{l-1}},\lambda_{l}) \Delta_l + \sqrt{\Delta_l\M{Q}(\lambda_{l})}\mathrm{d} \V{w}  \label{eq:EulerStochastic}
\vspace{1mm}
\end{equation}
where $\Delta_l \rmv=\rmv \lambda_l\rmv-\rmv\lambda_{l-1}$. In this way, particles $\{\V{x}_1^{(i)}\}_{i=1}^{N_\mathrm{s}} \! =\! \{\V{x}_{\lambda_{N_\lambda}}^{(i)}\}_{i=1}^{N_\mathrm{s}}$ representing the posterior \ac{pdf} $f(\V{x}|\V{z}) \propto \exp\big(\phi(\boldsymbol{x},\lambda\rmv=\rmv1)\big)$ are finally obtained.

Recent results demonstrate that, based on an appropriate choice of the diffusion term $\M{Q}(\lambda)$,  stochastic \ac{pfl} can provide a strongly reduced number of integration steps and computational complexity compared deterministic \ac{pfl} \cite{DaiDau:22}. A popular stochastic flow is Gromov's flow \cite{DuaHuaNou:18}, given by the drift
\begin{equation}\label{deterministicDrift}
  \V{\zeta}_{\mathrm{g}}(\V{x}, \lambda) = -\big(\nabla_{\rmv\V{x}}\nabla_{\rmv\V{x}}^\T \phi\big)^{-1} \nabla_{\rmv\V{x}} \log h
\end{equation}
and the diffusion matrix
\begin{equation}\label{diffusionMatrix}
\M{Q}_\mathrm{g}(\lambda) = -\big(\nabla_{\rmv\V{x}}\nabla_{\rmv\V{x}}^\T \phi\big)^{-1} \big(\nabla_{\rmv\V{x}}\nabla_{\rmv\V{x}}^\T \log h\big)\big(\nabla_{\rmv\V{x}}\nabla_{\rmv\V{x}}^\T \phi\big)^{-1}\rmv\rmv .
\end{equation}
Here, we used the short notation $\phi \triangleq \phi(\boldsymbol{x},\lambda)$ and $h \triangleq h(\V{x})$.

For a linear measurement model $\V{z} = \M{H} \V{x} + \V{v}$ with zero-mean additive Gaussian noise $\V{v}$ and covariance matrix $\M{R}$, we have $\nabla_{\rmv\V{x}} \log h = \M{H}^\T \M{R}^{-1}(\boldsymbol{z}-\M{H}\boldsymbol{x})$, $\nabla_{\rmv\V{x}}\nabla_{\rmv\V{x}}^\T \log h = -\M{H}^\T \M{R}^{-1}\M{H}$, and $\nabla_{\rmv\V{x}}\nabla_{\rmv\V{x}}^\T \phi = -\M{P}^{-1} - \lambda \M{H}^\T \M{R}^{-1}\M{H}$. We can now rewrite drift \eqref{deterministicDrift} and diffusion \eqref{diffusionMatrix} of Gromov's flow\vspace{1mm} as 
\begin{align}
  \V{\zeta}_{\mathrm{g}}(\V{x},\lambda) &= \big(\M{P}^{-1} + \lambda \M{H}^\T \M{R}^{-1}\M{H}\big)^{-1} \M{H}^\T \M{R}^{-1}(\boldsymbol{z}-\M{H}\boldsymbol{x}) \nn\\[0mm]
     \M{Q}_\mathrm{g}(\lambda)  &= (\M{P}^{-1} \rmv\rmv\rmv+\rmv\rmv \lambda \M{H}^\T \rmv\rmv\M{R}^{-1}\rmv\rmv\M{H})^{-1}\rmv(\M{H}^\T \rmv\rmv\M{R}^{-1}\rmv\rmv\M{H}) \nn\\[.5mm]
     &\hspace{22.5mm}\times \rmv\rmv (\M{P}^{-1} \rmv\rmv\rmv+\rmv\rmv \lambda \M{H}^\T\rmv\rmv \M{R}^{-1}\rmv\rmv\M{H})^{-1} \rmv. \nn \\[-4mm]
     \nn
\end{align}
For a nonlinear measurement model $\V{z} = \M{h}(\V{x}) + \V{v}$, one can linearize the measurement function at $\V{x} \rmv=\rmv \V{\mu}_{{l-1}}$ to obtain a linearized model. 
 The Gromov's flow applied to linearized measurement models has been demonstrated to outperform deterministic flows \cite{PalCoa:19, ZhaKhoMey:24, Cro:19}. However, as other deterministic and stochastic \acp{pfl}, due to approximations made for numerical integration, the Gromov's flow is not asymptotically optimal, i.e., the \ac{pdf} represented by the particles after the flow is only an approximation of $f(\V{x}|\V{z})$.



 An alternative approach for asymptotically optimal estimation is to use \ac{pfl} methods for \ac{is} within a particle filtering framework. Due to the lack of an invertible mapping in stochastic \ac{pfl}, evaluating the \ac{pdf} after the flow, as required for proposal evaluation \cite{LiCoates:17}, is typically infeasible\vspace*{-2mm}.
 

\section{\ac{is} With Stochastic \ac{pfl}}
\vspace{-1mm}

In this work, we propose to use \ac{pfl} based on a linearized measurement model to develop a ``flow-induced'' \ac{gmm} as proposal \ac{pdf}. For example, let us first transform the drift of the Gromov flow to an affine function, i.e,  $\V{\zeta}_{\mathrm{g}}(\boldsymbol{x},\lambda) = \M{A}_{\mathrm{g}}(\lambda)\boldsymbol{x} + \V{b}_{\mathrm{g}}(\lambda)$, \vspace{-2mm} with
\vspace{1mm}
\begin{align}
  \M{A}_{\mathrm{g}}(\lambda) &= -\Big(\M{P}^{-1} + \lambda \M{H}^\T \M{R}^{-1}\M{H}\Big)^{-1} \M{H}^\T \M{R}^{-1}\M{H} \nn\\[-.8mm]
  \V{b}_{\mathrm{g}}(\lambda) &= \Big(\M{P}^{-1} + \lambda \M{H}^\T \M{R}^{-1}\M{H}\Big)^{-1} \M{H}^\T \M{R}^{-1}\boldsymbol{z} \, .  \nn\\[-6mm] 
  \nn
\end{align}
Next, consider a single Gaussian $f(\V{x}) = \Set{N}(\V{x}; \V{\mu}_0,\M{P}_0)$ with mean ${\boldsymbol{\mu}}_0$ and covariance matrix $\M{P}_0$ as predicted/prior \ac{pdf}. Based on the affine form introduced above, we can migrate the mean and covariance of this Gaussian predicted/prior \ac{pdf} based on \ac{pfl}, i.e.\vspace{-.5mm},
\begin{align}
\V{\mu}_{l} &= \V{\mu}_{{l-1}} + \V{\zeta}_\mathrm{s}(\V{\mu}_{\lambda_{l-1}},\lambda_l)\Delta_l \label{meanTransform} \\[2mm]
\M{P}_{l} &= [\M{I}\rmv\rmv+\rmv\rmv\Delta_l\M{A}_{\mathrm{s}}\rmv(\lambda_l)]\M{P}_{{l\rmv-\rmv1}}[\M{I}\rmv\rmv+\rmv\rmv\Delta_l\M{A}_{\mathrm{s}}\rmv(\lambda_l)]^{\text{T}} + \Delta_l\M{Q}(\lambda_l) \label{covTransform}                                             
\end{align}
for $l = 1,\dots,N_\lambda$ and by setting $\V{\zeta}_\mathrm{s}(\V{\mu}_{\lambda_{l-1}},\lambda_l) \rmv=\rmv \V{\zeta}_\mathrm{g}(\V{\mu}_{\lambda_{l-1}},\lambda_l)$ and $\M{A}_{\mathrm{s}} \rmv=\rmv \M{A}_{\mathrm{g}}$. This principle can be extended to \acp{gmm} in a straightforward way. The resulting ``flow-induced''  \ac{gmm} can be used as a proposal \ac{pdf}, leading to an \ac{is} method that makes it possible to perform efficient nonlinear estimation in an asymptotically optimal manner.

For an accurate numerical implementation of the \ac{pfl}, the step sizes, $\Delta_l$, need to be adapted to the stiffness of the flow. A flow with reduced stiffness can be implemented with larger step sizes, i.e., fewer steps, and thus yields reduced computational complexity. Next, we investigate how to develop stochastic \acp{pfl} with reduced stiffness.
Recently, it has been shown that a solution to \eqref{eq:stochasticFlow} can be obtained by choosing $ \M{Q}(\lambda)$ arbitrarily and computing $\V{\zeta}_{\mathrm{s}}(\V{x}, \lambda)$ according to\vspace{-.5mm} \cite{DaiDau:22}
\begin{equation}
\V{\zeta}_{\mathrm{s}}(\V{x}, \lambda) = \V{\zeta}_{\mathrm{d}}(\V{x}, \lambda) + \frac{1}{2}  \ist \M{Q}(\lambda) \ist\ist \nabla_{\rmv\V{x}} \ist \phi(\boldsymbol{x},\lambda)
\label{stochasticFlowDev}
\end{equation}
where $\V{\zeta}_{\mathrm{d}}(\V{x}, \lambda)$ is a deterministic flow, i.e. a solution of the \ac{ode} $\mathrm{d} \V{x} = \V{\zeta}(\V{x}, \lambda) \mathrm{d}\lambda$. Based on this result, one can choose a deterministic \ac{pfl} and then design a diffusion matrix to get a stochastic \ac{pfl} with reduced stiffness. 

Consider the Gromov's flow with stochastic drift and diffusion as in \eqref{deterministicDrift} and \eqref{diffusionMatrix}. Based on \eqref{stochasticFlowDev}, the corresponding deterministic drift can be obtained\vspace{-.5mm} as 
\begin{equation}
\V{\zeta}_{\text{d-g}}(\V{x}, \lambda)  =  \V{\zeta}_{\mathrm{g}}(\V{x}, \lambda) -\frac{1}{2}\M{Q}_\mathrm{g}(\lambda) \ist\ist \nabla_{\rmv\V{x}} \ist \phi(\boldsymbol{x},\lambda).
\label{eq:deterministicGromov}
\end{equation}
By using an arbitrary diffusion matrix $\M{Q}(\lambda)$ and by substituting $\V{\zeta}_{\text{d}}(\V{x}, \lambda) $ in \eqref{stochasticFlowDev} by $\V{\zeta}_{\text{d-g}}(\V{x}, \lambda) $ in \eqref{eq:deterministicGromov}, a new stochastic drift can be developed\vspace{-.5mm} as
\begin{equation}
\V{\zeta}_{\mathrm{s}}(\V{x}, \lambda) = \V{\zeta}_{\mathrm{g}}(\V{x}, \lambda) + \frac{1}{2}  \ist (\M{Q}(\lambda)-\M{Q}_\mathrm{g}(\lambda))\ist\ist \nabla_{\rmv\V{x}} \ist \phi(\boldsymbol{x},\lambda)\, .
\label{stochasticFlowGromov}
\vspace{0mm}
\end{equation}

By considering a linearized measurement model, which results in $\nabla_{\rmv\V{x}}\phi(\boldsymbol{x},\lambda) = \M{P}^{-1}(\V{\mu}_0-\V{x})+\lambda \M{H}^\T \M{R}^{-1}(\boldsymbol{z}-\M{H}\boldsymbol{x})$, we can transform \eqref{stochasticFlowGromov} to an affine function, i.e., 
\begin{align}\label{eq:driftQA}
  \M{A}_\mathrm{s}(\lambda) = & -\frac{1}{2}\Big(\M{P}^{-1} + \lambda \M{H}^\T \M{R}^{-1}\M{H}\Big)^{-1} \M{H}^\T \M{R}^{-1}\M{H} \nn\\
                   &-\frac{1}{2}\M{Q}(\lambda)\big(\M{P}^{-1}+\lambda \M{H}^\T \M{R}^{-1}\M{H}\big) \\[-7.5mm]
                   \nn
\end{align}
\begin{align}
  \V{b}_\mathrm{s}(\lambda) = & \Big(\M{P}^{-1} + \lambda \M{H}^\T \M{R}^{-1}\M{H}\Big)^{-1} \M{H}^\T \M{R}^{-1}\boldsymbol{z} \nn\\[-.5mm]
                   & + \frac{1}{2}(\M{Q}(\lambda)-\M{Q}_\mathrm{g}(\lambda))\big(\M{P}^{-1}\V{\mu}_0+\lambda \M{H}^\T \M{R}^{-1}\V{z}\big)\, .  \nn\\[-4.5mm] 
                   \nn
\end{align}
Based on \eqref{meanTransform} and \eqref{covTransform}, this affine function can again be used to migrate the means and covariances of a \ac{gmm} representing a predicted/prior \ac{pdf} based on \ac{pfl}.

The eigenvalues of the diffusion matrixes $\M{Q}(\lambda)$ represent a tradeoff between the transient dynamics of the \ac{pfl} and numerical evaluation accuracy. The freedom to choose $\M{Q}(\lambda)$ makes it possible to optimize this tradeoff \cite{DaiDau:22}. In particular, the transient dynamics of the \ac{pfl} are measured by the condition number $\kappa(\cdot)$ of the nonsingular matrix $\M{A}_\mathrm{s}(\lambda)$ in \eqref{eq:driftQA}, i.e., the ratio of the largest singular value to the smallest singular value of nonsingular $\M{A}_\mathrm{s}(\lambda)$. A small condition number close to one usually implies a reduced stiffness of the \ac{sde} \cite{KloPla:B92}. Consider the following function form of $\M{Q}(\lambda)$ in \eqref{eq:driftQA}, i.e., $\M{Q} = c(\M{P}^{-1} + \lambda \M{H}^\T \M{R}^{-1}\M{H})^{-1}$ where $c$ is a constant. We can get $\lim_{c\rightarrow\infty}\kappa(\M{A}_{\mathrm{s}})=1$ for $\lim_{c\rightarrow\infty}||\M{Q}||=\infty$. However, we cannot make $||\M{Q}||$ too large, since it defines an upper bound of the numerical integration error using the Euler-Maruyama method \cite{DaiDau:22}. 

To balance the stiffness reduction and the error of numerical evaluation of the \ac{sde}, we adapt the objective function \cite{DaiDau:22}\vspace{0mm}  
\begin{equation}\label{objFunction}
    J(\M{Q}) = \kappa(\M{A}) + \alpha c
\end{equation}
where $\alpha$ is a hyperparameter.
The optimal solution $\M{Q}^\ast$ obtained by minimizing \eqref{objFunction} is $\M{Q}^\ast = c^\ast (\M{P}^{-1} + \lambda \M{H}^\T \M{R}^{-1}\M{H})^{-1}$\vspace{0mm}, with
\begin{equation}
c^\ast = \max\{ \sqrt{\frac{\overline{|\lambda|}-\underline{|\lambda|}}{\alpha}}-\underline{|\lambda|}, 0\}
\vspace{.5mm}
\end{equation}
where $\overline{|\lambda|}$ and $\underline{|\lambda|}$ are the max and min of all eigenvalues of the Jacobian matrix of $\V{\zeta}_\text{d-g}(\V{x}, \lambda)$ in \eqref{eq:deterministicGromov}, i.e., $\M{A}_\text{d-g} = -\frac{1}{2}(\M{P}^{-1} + \lambda \M{H}^\T \M{R}^{-1}\M{H})^{-1} \M{H}^\T \M{R}^{-1}\M{H}$. For a detailed proof, see \cite{DaiDau:22}. Recently, a simplified version of this solution has been introduced in~\cite{daum2024bullet}. This solution can offer higher computational efficiency in certain applications\vspace{-2mm}.

\section{Numerical Experiments and Results}

\label{sec:simRes}
We evaluate our flow-induced proposal \ac{pdf} in a 3-D source localization scenario where a volumetric array of receivers provides \ac{tdoa} measurements \cite{gustafsson03tdoa}. This scenario is complicated by (i) the highly nonlinear TDOA measurement model, (ii) measurement-origin uncertainty, and (iii) an unknown number of sources to be localized \cite{ZhaMey:21,ZhaMey:J24}. To address (ii) and (iii), we make use of the \ac{bp}-based message passing framework introduced in \cite{MeyKroWilLauHlaBraWin:J18,MeyBraWilHla:J17}. To address (i), we use our flow-induced Gaussian mixture proposal \ac{pdf} for weight computation in the belief update step and Monte Carlo integration in the message computation step. For more details on how \ac{pfl}-based proposal \ac{pdf} can be used within \ac{bp}-based message passing framework, see \cite{ZhaMey:21,ZhaMey:J24}\vspace{-2mm}.

\subsection{Source Localization Scenario and Implementation Aspects}
\begin{figure}[ht!]
  \centering
  \subcaptionbox{}{\hspace{-0mm}\includegraphics[scale=0.3]{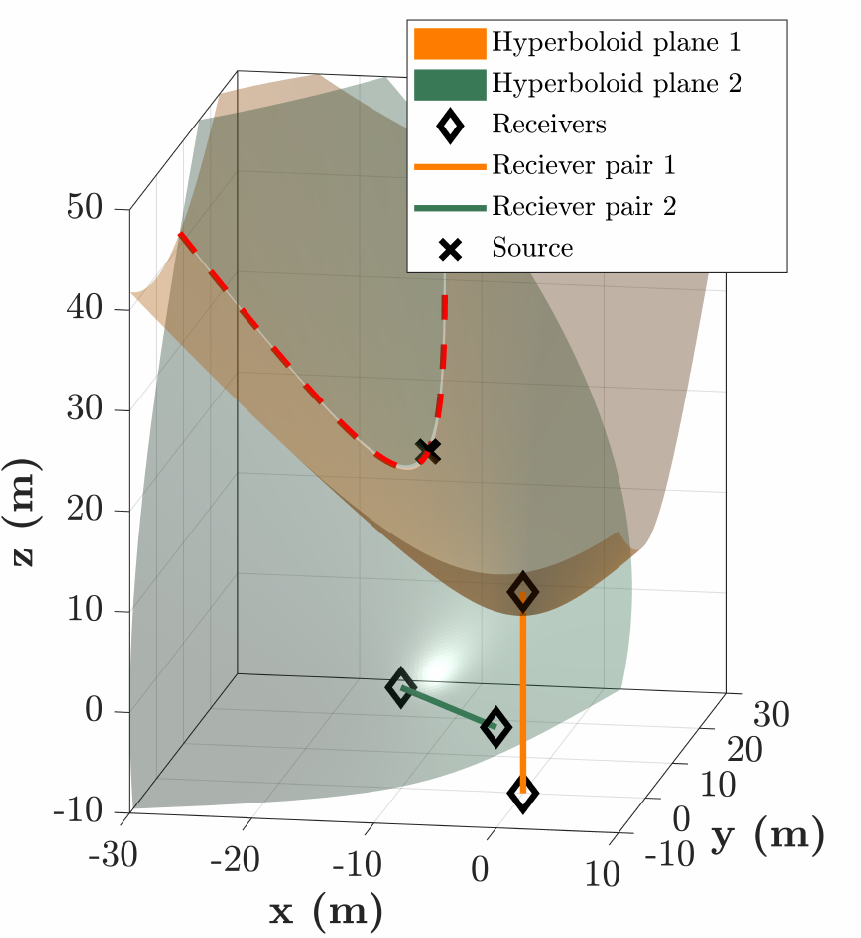}}
  \subcaptionbox{}{\includegraphics[scale=0.3]{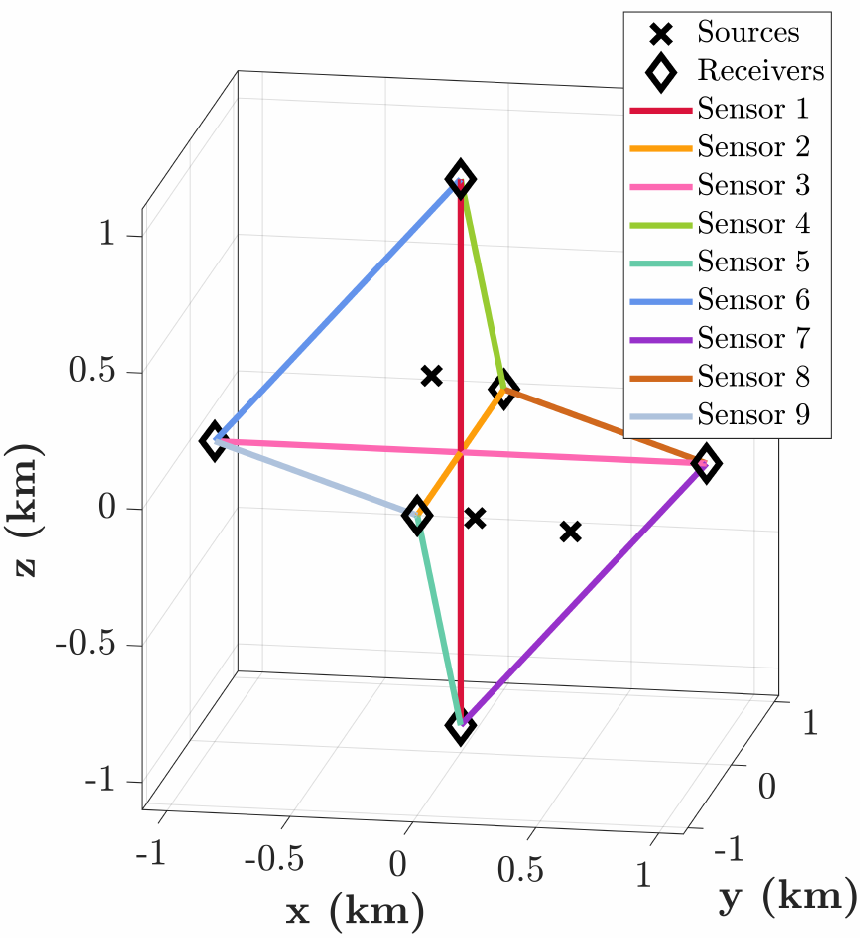}}
  \caption{(a). Source position and hyperboloids resulting from the \ac{tdoa} measurements of two sensors. Each sensor consists of two receiver pairs. A dashed red line indicates the intersection of the two hyperboloids. (b). Receiver and sensor topology used in our simulation. There are six receivers located at the center of each face of the \ac{roi} cube. Three sources are randomly placed in the \ac{roi}\vspace{-2mm}.}
  \label{fig:Hyperboloid}
  \vspace{-2mm}
\end{figure}

In this work, we consider the localization of an unknown number of static sources in a 3-D \ac{roi}. There are $V$ receivers. Pairs of receivers provide TDOA measurements, e.g., obtain by cross-correlation. In particular, the $m$-th TDOA measurements provided by the receiver with index $a$ and the receiver with index $b$, is modelled\vspace{1mm} as 
\begin{align}
z^{(m)}_{ab} &= \frac{1}{c} \Big(  \| \V{x}^{(j)}-\V{p}^{(a)} \|  - \| \V{x}^{(j)} - \V{p}^{(b)} \| \Big) + v^{(m)}_{ab} \nn \\[1mm]
&= h_{ab}(\V{x}^{(j)}) + v^{(m)}_{ab}. \label{eq:tau} \\[-4mm]
\nn
\end{align}
Here, $\V{p}^{(a)}$ and $\V{p}^{(a)}$ are the 3-D positions of the receivers, $c$ is the propagation speed in the considered medium, and $v^{(m)}_{ab}$ is the additive white noise with variance $\sigma_v^2$. The noise $v^{(m)}_{ab}$  is statistically independent across $m$ and across all receiver pairs $(a,b)$. The dependence of a measurement $z^{(m)}_{ab}$ on the source-location $\V{x}^{(j)}$ is described by the likelihood function $f(z^{(m)}_{ab}|\ist\V{x}^{(j)})$ that can be directly obtained from \eqref{eq:tau}. Note that, due to the nonlinear TDOA measurement model, this likelihood function has the shape of a hyperboloid (cf.~Fig.~\ref{fig:Hyperboloid}). For unambiguouse source localization, the measurements of multiple receiver pairs have to be used. Our scenario is further complicated by (ii) and (iii) discussed above.

Each receiver pair is considered one of $S$ sensors indexed by $s \in \{1,\dots,S\}$. The receivers of sensor $s$ are indexed $(s_a,s_b)$ and the number of measurements at sensor $s$ is $M_s$. We consider a topology with $V\rmv=\rmv6$ receivers and $S\rmv=\rmv9$ sensors is shown in Fig. \ref{fig:Hyperboloid}-b.
Furthermore, we set $\sigma_v = 0.5\text{ms}$ and $c=1500\text{m/s}$. The clutter measurements at sensor $s$ follow a uniform \ac{pdf} on $\|{\V{q}}^{(s_a)} - {\V{q}}^{(s_b)}\|/c$. The mean number of clutter measurements is $\mu_{\text{c}}=1$ and the probability of source detection, $p_{\text{d}}$, is set to $0.95$.  The \ac{roi} is defined as $[-1000\ist\text{m}, \ist 1000\ist\text{m} ]  \times [-1000\ist\text{m}, \ist 1000\ist\text{m}] \times [-1000\ist\text{m}, \ist 1000\ist\text{m}]$. 



Following \cite{MeyKroWilLauHlaBraWin:J18,ZhaMey:J24}, \ac{tdoa} measurements are processed sequentially across sensors. More precisely, let $f(\V{x}|\V{z}_{1:s-1})$ be the multimodal posterior \ac{pdf} after sensor update $s-1$.  This \ac{pdf} is represented by $N_{\mathrm{g}}$ Gaussian mixture components. For each component $k$ and each TDOA measurement, $z^{(s)}_m\rmv\rmv$, of sensor $s$, we perform \ac{pfl} and update each kernel mean and covariance matrix based on \eqref{meanTransform} and \eqref{covTransform}. The resulting $N_\mathrm{k} \times M_s$  Gaussian components will be used for \ac{is} and Monte Carlo integration within \ac{bp}-based message passing \cite{MeyKroWilLauHlaBraWin:J18}. The result is an approximate posterior \acp{pdf} $f(\V{x}|\V{z}_{1:s})$ represented by $N_{\mathrm{g}}$ Gaussian mixture components. For more details about \ac{bp} message passing with \ac{pfl}, see \cite{ZhaMey:J24,ZhaKhoMey:24}.


As a reference methods for the proposed \ac{is} with \textit{optimized stochastic} \ac{pfl} (``\ac{pfl}-OS"), we consider bootstrap \ac{is} (``BS"), which directly uses the posterior \ac{pdf} $f(\V{x}|\V{z}_{1:s-1})$ from previous sensor $s\rmv-\rmv1$ as proposal \ac{pdf}. We also use methods with ``flow-induced'' proposal \acp{pdf} using the \textit{deterministic} \ac{edh} \ac{pfl} (``\ac{pfl}-D") \cite{ZhaMey:J24}) and the \textit{stochastic} Gromov's \ac{pfl} (``\ac{pfl}-S") \cite{ZhaKhoMey:24}). Since every method yields different stiffness, for numerical integration, every method requires a different resolution of the pseudo-time. This resolution is defined as the inverse of the time-interval $\Delta_l$. In addition, for every method, a higher resolution is typically needed at the first few steps of the numerical integration. We use an exponentially increasing ratio of the time-interval, i.e., $\Delta_l = \beta \Delta_{l-1}$, where $\beta$ is a fixed ratio for $l=2,\dots,N_\lambda-1$. Then, we control the interval by two parameters, the initial difference $\Delta_1$ and the increasing ratio $\beta$. Note that a larger $\Delta_1$ and $\beta$ result in fewer discrete steps $N_\lambda$ and thus to a reduced runtime.  For each method, we choose $\Delta_1$ and $\beta$ to obtain a good runtime-accuracy tradeoff \vspace{-4.5mm}.

\subsection{Results}
Table \ref{table1} shows the mean \ac{ospa} error \cite{SchVoVo:J08} (with a cutoff threshold at 30) and runtime per run for the different methods and different parameters settings. OSPA and runtime are averaged over 100 Monte Carlo runs.  We also list the number of Gaussian components, $N_\mathrm{g}$, as well as the number of samples per component, $N_\mathrm{p}$. 
\begin{table}[ht!]
\small
  \centering 
  \renewcommand{\arraystretch}{1.25}
  \begin{tabular}{ C{0.3cm} | C{1.1cm} | C{1.3cm} | C{1.5cm} | C{0.8cm} | C{1.3cm} }
  \hline
  ID & Method & $(N_\mathrm{g},N_\mathrm{p})$ & $(\beta,\Delta_1)$ & OSPA & Runtime(s)\\[.3mm]
  \hline
  1 & BS & $(-,2\text{e}6)$ & $(-,-)$ & $25.89$ & $47.0$ \\
  2 & BS & $(-,4\text{e}7)$ & $(-,-)$ & $14.73$ & $718.8$ \\[-.3mm]
  3 & \ac{pfl}-D & $(100,5\text{e}3)$ & $(1.5,1\text{e}\rmv\rmv-\rmv\rmv7)$ & $11.1$ & $202.2$ \\[-.3mm]
  4 & \ac{pfl}-S & $(100,5\text{e}3)$ & $(1.5,1\text{e}\rmv\rmv-\rmv\rmv7)$ & $8.36$ & $612.7$ \\[-.3mm]
  5 & \ac{pfl}-OS & $(100,5\text{e}3)$ & $(1.5,1\text{e}\rmv\rmv-\rmv\rmv5)$ & $6.33$ & $490.9$ \\[-.3mm]
  6 & \ac{pfl}-OS & $(100,5\text{e}3)$ & $(2,1\text{e}\rmv\rmv-\rmv\rmv4)$ & $6.70$ & $295.2$ \\[-.3mm]
  \hline
  \end{tabular}
  \vspace{0mm}
  \caption{Simulated mean \ac{ospa} error and runtime per run for different methods and system parameters.}\label{table1}
  \vspace{-2mm}
\end{table}
\vspace{3mm}

BS suffers from particle degeneracy and thus yields the highest OSPA. In addition, the large number of $4\text{e}7$ particles results in the largest memory requirements. In contrast, \ac{pfl}-based methods required much fewer particles. It can be seen that, for the same $N_\lambda$ value, \ac{pfl}-S has a smaller \ac{ospa} error compared to \ac{pfl}-D while its runtime is three times larger. \ac{pfl}-OS relies on a diffusion matrix $\M{Q}$ that has been minimized based on \eqref{objFunction} by setting $\alpha=0.1$. Notably, \ac{pfl}-OS yields the lowest \ac{ospa} error and, at the same time, has a low runtime.

To better understand the influence of $\alpha$ on the estimation error for different step sizes, we compare the proposed methods for different values of $\alpha$ and different resolutions of pseudo time. Results are shown in Table \ref{table2}.  \ac{pfl}-D and \ac{pfl}-S results are also listed. When the step size is small, \ac{pfl} can lead to accurate results despite the stiffness of the underlying \acp{ode} and \acp{sde}. In this case, traditional methods such as \ac{edh} and Gromov's flow also perform well. However, the small step size comes at the cost of a strongly increased runtime. As the step size increases, only \ac{pfl}-OS with carefully chosen parameter $\alpha$ (cf. \eqref{objFunction}) results in a good estimation accuracy. Importantly, our results indicate that \ac{pfl}-OS has a significantly improved complexity accuracy tradeoff compared to \ac{pfl}-D and \ac{pfl}-S. Further numerical analysis is provided in \cite{ZhaKhoAta:J24SM}. Here, the tradeoff between the condition number of the Jacobian matrix $\M{A}$ and the norm of the diffusion matrix $\M{Q}$ is analyzed numerically for different values of $\alpha$\vspace{-1mm}.
\begin{table}[ht!]
  \centering 
  \small
\renewcommand{\arraystretch}{1.25}
\begin{tabular}{|@{}c|c|c|c|}
\hline
\diagbox[width=9em,trim=l,innerleftsep=.3cm,innerrightsep=0.2cm]{Method}{$(\beta,\Delta_1)$} & $(1.3,1\text{e}\rmv\rmv-\rmv\rmv13)$ & $(1.5,1\text{e}\rmv\rmv-\rmv\rmv5)$ & $(2,1\text{e}\rmv\rmv-\rmv\rmv4)$\\
\hline
\ac{pfl}-D & \cellcolor{blue!15}6.29 & 30 & 30\\
\hline
\ac{pfl}-S &  8.70 & 24.51 & 28.4 \\
\hline
\ac{pfl}-OS ($\alpha=0.01$) &  \cellcolor{blue!35}4.79 & 30 & 30 \\
\hline
\ac{pfl}-OS ($\alpha=0.1$) &  10.28 & \cellcolor{blue!35}6.33 & \cellcolor{blue!35}6.70 \\
\hline
\ac{pfl}-OS ($\alpha=0.5$) &  10.32 & \cellcolor{blue!15}9.36 & \cellcolor{blue!15}12.85 \\
\hline
\end{tabular}
  \vspace{0mm}
  \caption{Mean \ac{ospa} error of different \ac{pfl}-based \ac{is} method and different integration step sizes.}\label{table2}
  \vspace{-5mm}
\end{table}

\section{Conclusion}
\vspace{-.5mm}

In this paper,  we introduced a general approach to perform \ac{is} based on stochastic \ac{pfl}. Stochastic \ac{pfl} introduces a diffusion term in the \ac{ode} that describes particle motion. A carefully determined diffusion term reduces the stiffness of the ODE and makes it possible to perform \ac{pfl} with a lower number of numerical integration steps compared to traditional deterministic \ac{pfl}. Our method makes it possible to evaluate a ``flow-induced'' proposal \ac{pdf} after the parameters of a \ac{gmm} have been migrated by stochastic \ac{pfl}. Compared to conventional stochastic \ac{pfl}, the resulting updating step is asymptotically optimal. Within our method, it is possible to optimize the diffusion matrix that describes the diffusion term of the ODE to improve the accuracy-computational complexity tradeoff. The presented numerical results in a highly nonlinear 3-D source localization scenario showcased a reduced stiffness of the ODE and an improved estimating accuracy compared to state-of-the-art deterministic and stochastic \ac{pfl}. Further research includes flow-induced \ac{is} for different types of measurement models \cite{MeyWil:J21}, application of flow-induced \ac{is} to real-world problems \cite{LiaMey:J24,WatStiTes:C24}, and efficient information-seeking control methods based on \ac{pfl} \cite{ZhaTeaMey:22,PlaStrCar:J23,AsgAta:J23}.

\bibliographystyle{IEEEtran}
\bibliography{StringDefinitions,IEEEabrv,references,Books}




\end{document}